\documentclass[journal,twoside]{IEEEtran}

\usepackage{amsmath,amssymb,amsfonts}
\usepackage{algorithmic}
\usepackage{hyperref}
\usepackage{graphicx}
\usepackage{textcomp}
\usepackage{biblatex}
\usepackage{multirow}

\begin{document}
\title{TEAM PILOT - Learned Feasible Extendable Set of Dynamic MRI Acquisition Trajectories}
\author{Tamir Shor$^{1}$, Chaim Baskin$^{2}$, Alex Bronstein$^{1}$
\thanks{}
\thanks{$^{1}$Department of Computer Science, Technion Institute of Technology, Haifa, Israel.}
\thanks{$^{2}$School of Electrical and Computer Engineering, Ben-Gurion University of The Negev, Be'er Sheva, Israel.}
\thanks{Corresponding Author Tamir Shor: tamir.shor@campus.technion.ac.il}
}

\maketitle

\begin{abstract}
Dynamic Magnetic Resonance Imaging (MRI) is a crucial non-invasive method used to capture the movement of internal organs and tissues, making it a key tool for medical diagnosis. However, dynamic MRI faces a major challenge: long acquisition times needed to achieve high spatial and temporal resolution. This leads to higher costs, patient discomfort, motion artifacts, and lower image quality.
Compressed Sensing (CS) addresses this problem by acquiring a reduced amount of MR data in the Fourier domain, based on a chosen sampling pattern, and reconstructing the full image from this partial data. While various deep learning methods have been developed to optimize these sampling patterns and improve reconstruction, they often struggle with slow optimization and inference times or are limited to specific temporal dimensions used during training.
In this work, we introduce a novel deep-compressed sensing approach that uses 3D window attention and flexible, temporally extendable acquisition trajectories. Our method significantly reduces both training and inference times compared to existing approaches, while also adapting to different temporal dimensions during inference without requiring additional training. Tests with real data show that our approach outperforms current state-of-the-art techniques. The code for reproducing all experiments will be made available upon acceptance of the paper.
\end{abstract}

\begin{IEEEkeywords}
Magnetic Resonance Imaging (MRI), fast image acquisition, image reconstruction, dynamic MRI, deep learning.
\end{IEEEkeywords}

\section{Introduction}
\label{sec:introduction}

Magnetic Resonance Imaging (MRI) has long been a preferred medical imaging technique due to its ability to provide high-quality soft-tissue contrast in a non-invasive way without exposing patients to harmful radiation. Dynamic MRI, which captures multiple frames over time, is particularly useful in applications such as cardiac imaging, tissue motion analysis, and cerebrospinal fluid (CSF) flow studies, where static MRI falls short \cite{alperin1996hemodynamically,michelini2018dynamic,zhang1999intrahepatic}.

A significant drawback of MRI, however, is the long scan times needed to obtain accurate, high-resolution images. This issue becomes even more pronounced in dynamic MRI, where both high spatial and temporal resolution are required. Prolonged scan times not only lead to increased patient discomfort and higher costs but also demand that patients remain still for longer periods, increasing the likelihood of motion artifacts that degrade image quality. These challenges have sparked growing research interest in reducing MRI scan times.

One popular approach to address this is Compressed Sensing (CS), which reduces scan time by subsampling the image’s $k$-space using a predetermined trajectory \cite{lustig2007sparse}. CS techniques are applied before reconstruction methods that aim to recover lost information from subsampling and filter out blurring and aliasing artifacts caused by undersampling below the Nyquist criterion\cite{zaitsev2015motion}. 

Previous studies \cite{wang2021b,weiss2019pilot,shor2023multi} have demonstrated that the best results in CS are obtained by learning the acquisition trajectories for subsampling simultaneously with the reconstruction network. However, this joint optimization is challenging because the optimization of one component (e.g., the reconstruction network) directly influences the input and gradients of the other (e.g., acquisition trajectories), leading to potential inefficiencies during training. Additionally, learning acquisition trajectories must take into account the kinematic constraints of the MRI machine. In the dynamic MRI context, these challenges are even more critical, as models must identify and differentiate between features across the temporal dimension to avoid sampling redundant information across frames.

The current state-of-the-art in dynamic CS is the Multi-PILOT\cite{shor2023multi} which jointly optimizes non-Cartesian $k$-space acquisition trajectories for each frame alongside a reconstruction network. While it achieves impressive results in temporal MRI reconstruction, Multi-PILOT suffers from prolonged optimization times that scale linearly with the temporal dimension. Moreover, it lacks generalization across different temporal dimensions—a model trained on $8$ frames performs poorly when extended to $16$ frames.

In this work, we introduce \textbf{T}emporally \textbf{E}xtendible \textbf{A}ttention-based \textbf{M}ulti PILOT (TEAM-PILOT). TEAM-PILOT builds on the Multi-PILOT framework but introduces modifications to both the reconstruction network and the trajectory learning process to address its predecessor's limitations. We demonstrate that TEAM-PILOT not only improves reconstruction performance but also significantly reduces training time. Additionally, our method generalizes across different temporal dimensions, allowing it to handle any number of frames during inference, regardless of the number of frames used during training.

\section{Related Work}
Lustig's seminal paper \cite{lustig2007sparse} was probably the first to demonstrate the potential of sparse sampling for the acceleration of MRI acquisition. Since then, numerous studies have proposed various techniques to optimize a set of feasible sampling points for compressed sensing.
Following the immense progress in the development of deep learning-based tools and their potential in inverse problem solving \cite{Ongie:IEEE:2020}, the vast majority of recent research focuses on modeling the sparse sampling problem as an inverse problem, where a deep neural model is used to reconstruct the fully-sampled signal given the downsampled input.
One approach for sparse sampling focuses on establishing some constant set of handcrafted acquisition trajectories (e.g., Cartesian \cite{weiss2020joint}, Radial \cite{bilgin2008randomly}, and Golden Angle \cite{zhou2017golden}  -- henceforth collectively referred to as \emph{fixed trajectories} in this paper). Sub-Nyquist sampling of the $k$-space is performed using these trajectories and deep learning models are subsequently applied to restore the image data lost in undersampling \cite{hammernik2018learning, hyun2018deep}, or to perform super-resolution reconstruction \cite{chen2020mri,masutani2020deep}. This approach has been popular both in the context of static \cite{larson2007anisotropic,yiasemis2023retrospective} and dynamic \cite{utzschneider2021towards,bliesener2020impact} MR imaging. 

While this approach is, at least conceptually, simpler to model, implement, and optimize, current research also explores the modeling and optimization of the acquisition trajectories themselves, in a differentiable and physically feasible manner.
While trajectory optimization can also be performed over a set of Cartesian subsampling schemes (namely, trajectories that lie on a Cartesian grid) \cite{weiss2020joint, bahadir2020deep}, recent research \cite{shor2023multi} showed that non-Cartesian learning of the acquisition trajectories significantly surpasses all other methods in both static \cite{weiss2019pilot} and dynamic \cite{shor2023multi} CS.

As previously mentioned, while effective, non-Cartesian trajectory learning poses significant optimization challenges -- the trajectory parameters and the reconstruction network strongly affect each other and are both constantly changing during the training, making it unstable. Furthermore, such optimization requires modeling and injecting into the optimization scheme a set of hardware-related kinematic constraints on the trajectory. Lastly, in the dynamic setting, efficient learning must make use of the data acquired across the temporal frames (e.g., avoid sampling similar, irrelevant, or temporally static data several times across different frames). The solution proposed in the current state-of-the-art approach, Multi-PILOT (\cite{shor2023multi}, addresses these challenges by using 2D attention for the reconstruction network alongside training techniques such as resetting the reconstruction network parameters every several epochs and optimizing trajectory parameters separately across temporal frames. While efficient, this solution suffers from several limitations: Firstly, it is computationally intensive -- the need to optimize every frame separately requires several days of optimization on a modern GPU to achieve the reported performance. Secondly, optimization time complexity grows linearly with the number of frames. Lastly, this solution does not generalize across temporal dimensions in the sense that a different number of temporal frames requires full training from the beginning. 

We speculate that these difficulties mainly originate from the inefficient cross-frame relationships learned via spatial (2D) attention, and therefore in this study, we propose a novel, spatio-temporal (3D) attention-based pipeline. We show that our algorithm achieves superior reconstruction results, while greatly alleviating the mentioned difficulties found in previous methods.

\section{Method}
\begin{figure*}[htbp]

\begin{center}

  \includegraphics[width=\linewidth]{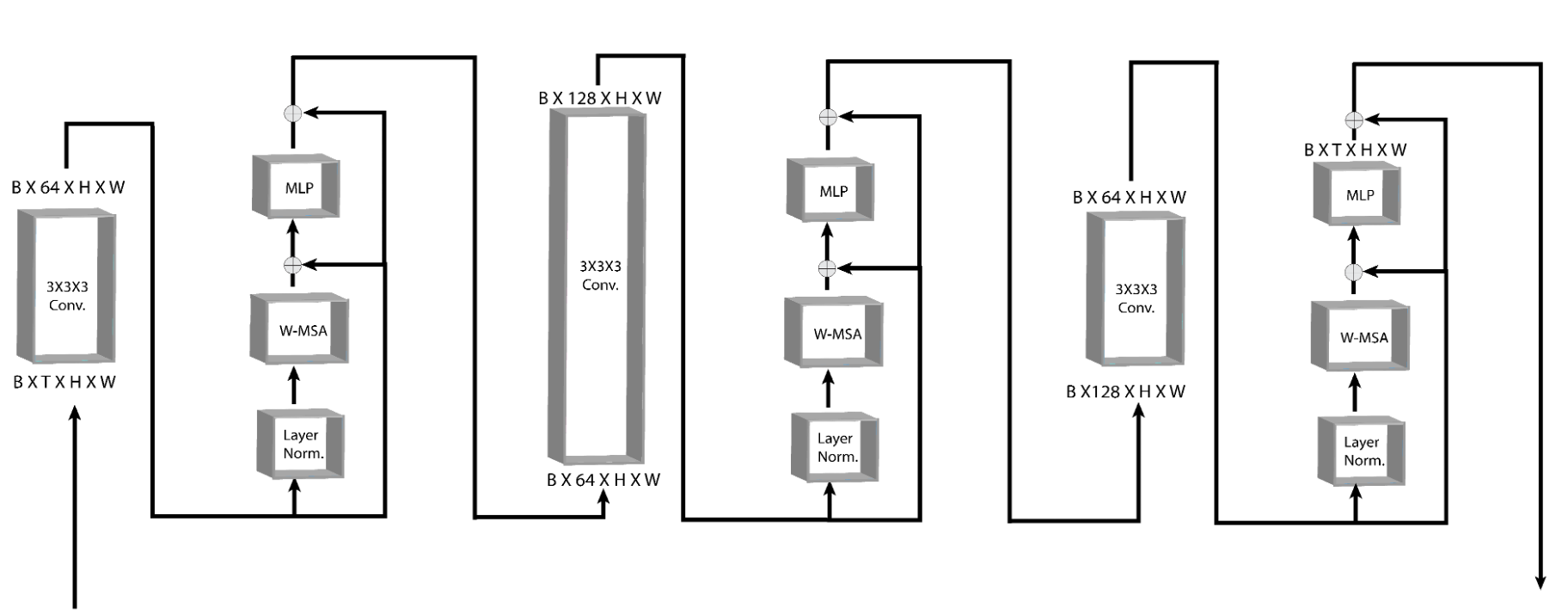}
  \end{center}
  {\caption{\textbf{3D attention-based reconstruction model.} {The model receives the downsampled frames $\mathbf{\tilde{Z}}$ in the image domain as the input, processes them using a combination of 3D convolution and windowed attention blocks, outputting the reconstructed frames $\mathbf{\hat{Z}}$.} }
  \label{fig:model}
  }
\end{figure*}

\begin{figure*}[htbp]

\begin{center}

  \includegraphics[width=0.8\linewidth]{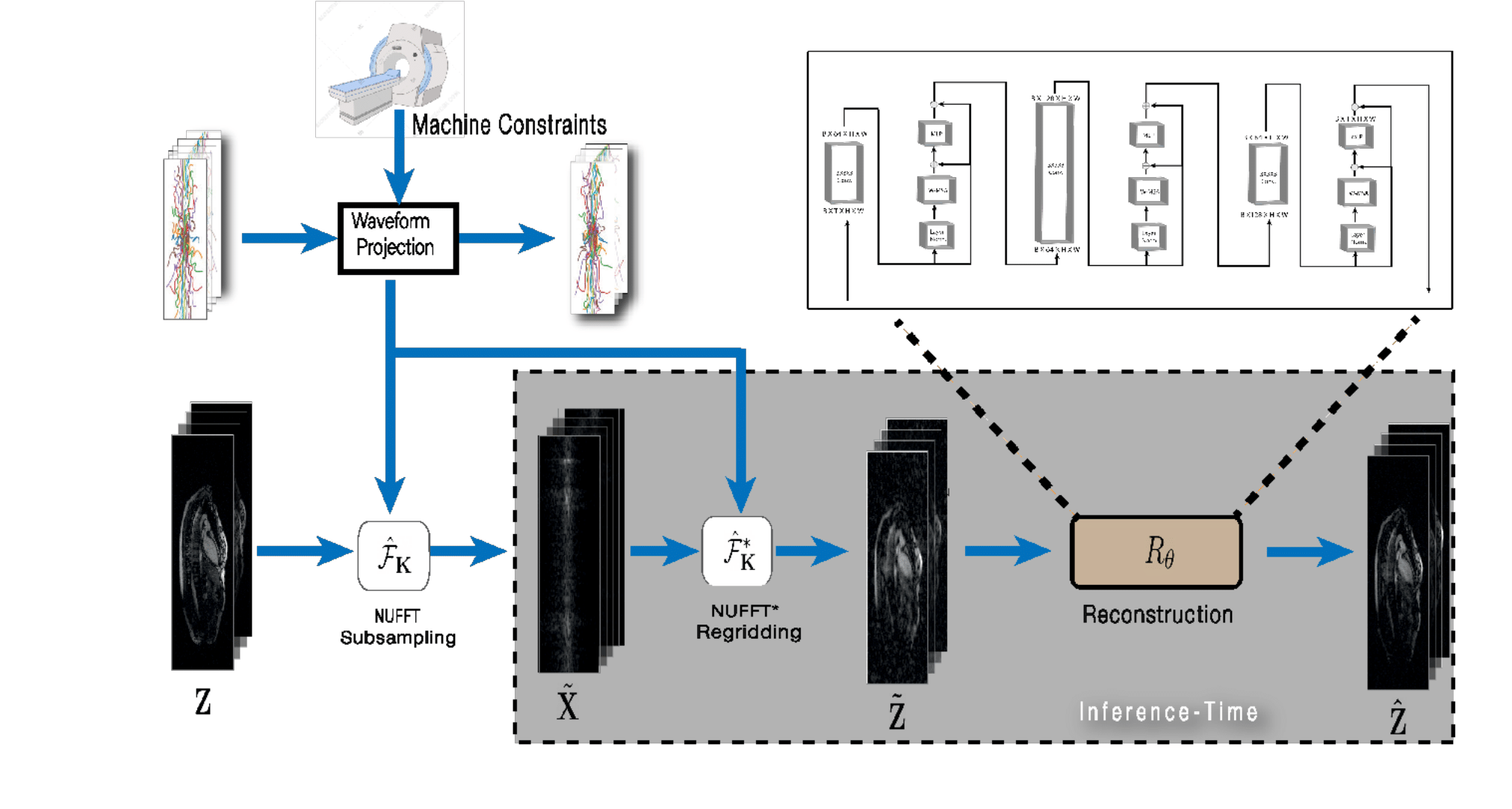}
  \end{center}
  {\caption{\textbf{Full data processing pipeline} {including the emulation of data acquisition and image reconstruction. Fully sampled frames $\mathbf{Z}$ serving as the ``ground-truth" are fed into the pipeline alongside with the acquisition trajectories $\mathbf{K}$. The reconstructed frames are received at the output.} }
   \label{fig:pipeline}
  }
 
\end{figure*}

In the following section, we introduce the model architecture and optimization strategies utilized in TEAM-PILOT to address the challenges encountered by previous methods, as discussed earlier. Section \ref{subsmodel} outlines our parameterized Compressed Sensing pipeline, while section \ref{period} presents our novel optimization technique designed to enhance temporal generalizability in trajectory stacking.
\subsection{Model} 
\label{subsmodel} 
We build upon the framework used in PILOT and Multi-PILOT, as proposed by \cite{weiss2019pilot,shor2023multi}. The process begins with a subsampling layer that parameterizes trajectory acquisition and subsampling. Next, a regridding layer maps the subsampled image onto the Cartesian grid, followed by a reconstruction layer that recovers full data from the subsampled data in the image domain. The parameters learned in this model include the set of acquisition trajectories $\mathbf{K}$ and the reconstruction model parameters $\theta$. Our primary enhancement to the Multi-PILOT architecture in this work is the network employed to model the reconstruction operator $R_{\boldsymbol{\theta}}$. The following sections provide a detailed explanation of each stage in the pipeline.

\subsubsection{Subsampling layer}
\label{subs}
This layer models the subsampling operator, which is determined by the current parameters of the $k$-space acquisition trajectories $\mathbf{K}\in\mathbb{R}^{N_\textrm{frames}\times N_\textrm{shots}\times m}$. Similar to Multi-PILOT, we maintain the simplifying assumption that dynamic MR images are sampled as a discrete set of temporal frames. Here, $N_\textrm{frames}$ represents a hyperparameter that specifies the number of learned acquisition trajectories we model, $N_\textrm{shots}$ is the number of RF excitations per frame, and $m$ is the number of acquisition points sampled during each RF excitation. The input to this layer consists of fully-sampled $k$-space data for $n$ temporally successive frames, denoted as $\mathbf{Z}=\left(\mathbf{Z}_1, \mathbf{Z}_2, \dots, \mathbf{Z}_n\right) \in \mathbb{C}^{n\times H\times W}$, where $W$ and $H$ refer to the width and height dimensions, respectively.
Since we aim to allow non-Cartesian acquisition trajectories, uniform spacing between acquisition points cannot be guaranteed, making it necessary to apply the non-uniform fast Fourier transform (NUFFT) algorithm \cite{dutt1993fast} to obtain the downsampled image in the frequency domain. Additionally, to ensure that the learned acquisition trajectories are physically feasible, we must account for machine-related kinematic constraints. To achieve this, we project $\mathbf{K}$ onto the kinematically feasible set using the algorithm from \cite{chauffert2016projection}.
The output of this layer is a set $\tilde{\mathbf{X}}=\hat{\mathcal{F}}_\mathbf{K}(\mathbf{Z}) \in \mathbb{C}^{n\times H\times W}$, representing the $n$ subsampled frames in the frequency domain.

\subsubsection{Regridding Layer} 
This layer takes as input the $n$ frames of the subsampled $k$-space data, $\tilde{\mathbf{X}}$, from the subsampling layer and applies the adjoint (inverse) NUFFT operator \cite{dutt1993fast} to convert the subsampled $k$-space data into $n$ subsampled frames in the image domain, $\tilde{\mathbf{Z}} = \hat{\mathcal{F}}^{*}_{\mathbf{K}} (\tilde{\mathbf{X}}) \in \mathbb{R}^{n\times H\times W}$. As a result, the subsampled $k$-space data are mapped onto a Cartesian grid in the image domain. This transformation is achieved by the NUFFT algorithm, which first performs resampling and interpolation operations, followed by a standard FFT to generate the final output on the desired grid. Full details on the differentiability of these operations, which is essential for the learning process, can be found in \cite{weiss2019pilot}.

\subsubsection{Reconstruction Layer} 
As mentioned earlier, subsampling $k$-space data leads to violations of the Nyquist criterion, resulting in artifacts in the reconstructed image \cite{zaitsev2015motion}, often manifesting as complex patterns that are difficult to eliminate. One of the latest techniques for removing aliasing artifacts involves the use of a denoising neural network as the reconstruction model. In our framework, this reconstruction model takes the downsampled frames $\mathbf{\tilde{Z}}$ in the image domain and outputs a set of reconstructed frames, $\mathbf{\hat{{Z}}} = R_{\boldsymbol{\theta}}(\mathbf{\tilde{Z}})$, where $R_{\boldsymbol{\theta}}$ represents the reconstruction network parameterized by ${\boldsymbol{\theta}}$. The network is trained by minimizing the following objective: \begin{equation} \min_{\mathbf{K}, \boldsymbol{\theta}} , \sum_{i} \mathcal{L}(R_{\boldsymbol{\theta}}(\hat{\mathcal{F}}^{*}{\mathbf{K}}(\hat{\mathcal{F}}{\mathbf{K}} (\mathbf{Z}_i))), \mathbf{Z}_i), \label{eq} \end{equation} jointly with respect to the network parameters and the acquisition trajectories $\mathbf{K}$.

Our primary modification to the Multi-PILOT architecture is within the reconstruction layer. The extended optimization times in Multi-PILOT are mainly due to the requirement of performing per-frame optimization before combining extracted features across the temporal domain. To overcome this limitation, we propose replacing this mechanism with a 3D attention-guided model, facilitating more efficient feature learning across frames. However, basic 3D attention attends every image patch to every other patch, which may not yield the desired decrease in optimization time and resource usage due to the large number of patches involved in attention computations. Therefore, we adopt the window multi-headed shifted attention (W-MSA) blocks proposed in the Video Swin Transformer paper \cite{liu2022video}. This attention mechanism performs cross-patch attention within certain window localities that are shifted across blocks, thus reducing computational costs while potentially allowing each pair of patches to attend to one another.

While utilizing the entire Video Swin Transformer architecture is applicable to our case, we empirically found that using 3D convolution for feature extraction prior to the unshifted window attention layers was more beneficial than employing the patch embedding layers proposed in the original architecture. We also observed that in the proposed architecture, attention shifting increased computation times without significantly enhancing reconstruction results. Therefore, we perform only unshifted window attention. We attribute this finding to convolutional feature extraction enabling "communication" between patches from different windows, thereby partially replacing the effect of shifting. An overview of our reconstruction network is depicted in Figure \ref{fig:model}. The full acquisition and reconstruction pipeline is presented in Figure \ref{fig:pipeline}.


\subsection{Optimization}
\label{period}
In Multi-PILOT, the dataset is divided into units of size $k \times H \times W$, where $k$ represents a fixed temporal duration. A model trained on sequences of length $k$ can then be applied to sequences of arbitrary length during inference by performing simple trajectory stacking. This involves partitioning the sequence into units of length $k$ and padding the remainder before performing serial inference on data segments with a temporal dimension of $k$. For instance, a model trained on a sequence of length $k=8$ can be applied to reconstruct a sequence of length $k=27$ by performing inference on frames 1–8, 8–16, 16–24, and 24–27 (with padding applied to the final segment to maintain a dimension of 8 frames). However, this approach introduces jittering and artifacts between frames from consecutive partitions of length $k$, as further demonstrated in Section \ref{perres}. This limits Multi-PILOT's generalizability to sequences of varying temporal lengths.

To address this limitation, we propose several epochs of post-training trajectory refinement, where training is performed on data partitioned into sequences of $2k$ frames, rather than $k$. Model evaluation is conducted using simple sequence stacking. During this refinement stage, we extend the optimization criterion from equation (\ref{eq}) by adding a regularization term to encourage smoother transitions between consecutive sequences of length $k$. This regularization penalizes the mean temporal derivative of the stacked output (temporal length $2k$) relative to the mean temporal derivative of the non-stacked output (temporal length $k$), where no artifacts are present.

Given a data sample $x \in \mathbb{R}^{k \times H \times W}$, we define the mean temporal derivative vector $\mu_k \in \mathbb{R}^{k-1}$ element-wise as $\mu_k(l) = \frac{1}{H \cdot W} \sum_{i=0}^{H} \sum_{j=0}^{W} \left( x[t+1, i, j] - x[t, i, j] \right)$, for all $0 \leq t \leq k-1$.

Our goal is to penalize the discrepancy between the temporal derivatives of stacked ($2k$-length) and non-stacked ($k$-length) sequences. To estimate the expected temporal derivative values, we use simple averaging of derivatives across the training set. As a preliminary step, before the trajectory refinement stage (when data is partitioned into sequences of length $k$), we compute $\mu_k$ for every such sequence and calculate the average across all entries: $\tilde{\mu_k} \triangleq \frac{1}{k-1} \sum_{t=0}^{k-1} \mu_k(t) \in \mathbb{R}$. We then compute the average value of $\tilde{\mu_k}$ across all training samples, denoted as $\mu_\mathcal{X}$. This scalar represents the characteristic value of the temporal derivative when trajectory stacking does not induce abnormal behavior, and it is stored for use in trajectory refinement.

During trajectory refinement, for each sample (of sequence length $2k$), we compute $\mu_{2k} \in \mathbb{R}^{2k \times W \times H}$. These values allow us to penalize abnormal temporal derivatives, which is incorporated into the following modified optimization criterion used during the refinement stage:


\begin{equation}
\begin{split}
\min_{\boldsymbol{K}, \boldsymbol{\theta}} \, \sum_{i} \mathcal{L}(R_{\boldsymbol{\theta}}(\hat{\mathcal{F}}^{*}_{\boldsymbol{K}}(\hat{\mathcal{F}}_{\boldsymbol{K}} (\boldsymbol{Z}_i))), \boldsymbol{Z}_i) + \\ \lambda_{ref}\cdot\sum_{t=0}^{2k-1}\max\{(\mu_{2k}(t)-\mu_{\mathcal{X}}), 0\}
\end{split}
\label{eq:min2}
\end{equation}

The term $\lambda_{ref}$ represents the regularization weighting factor. By incorporating this regularization, we achieve smoother transitions between consecutive frames from different partitions. As demonstrated in Section \ref{perres}, this leads to improved generalization for sequences of varying lengths without significantly affecting the reconstruction performance compared to the non-regularized optimization criteria.

\section{Results}
\label{exps}
In the following section, we present the experimental setup for our proposed method. Sections \ref{data} and \ref{setup} describe the data used and the optimization settings. In Section \ref{compres}, we conduct an ablation study comparing TEAM-PILOT's performance to other learned and non-learned acquisition scheme baselines. Section \ref{acqres} highlights our method's ability to reduce acquisition time compared to the current state-of-the-art, and Section \ref{attres} further demonstrates the advantages of our reconstruction model (Figure \ref{fig:model}).
\subsection{Data}
\label{data}
We used the augmented version of the OCMR dataset \cite{chen2020ocmr}, following the augmentation procedures described in \cite{shor2023multi}. This dataset consists of 265 anonymized cardiovascular MRI (CMR) scans, including both fully sampled and undersampled multi-coil data, which were augmented into 4,170 CINE MRI videos, each containing units of 8 temporal frames.

\subsection{Experimental setup}
\label{setup}
All experiments were conducted on a single NVIDIA RTX-2080 GPU. In each experiment, optimization was performed over 170 epochs for the primary optimization stage, followed by 10 additional epochs for the trajectory refinement stage. The Adam optimizer \cite{kingma2014adam} was used, with a learning rate of 0.05 for the trajectory parameters and $10^{-4}$ for the reconstruction model parameters. For all experiments, we utilized batches of 12 samples, each containing eight $384 \times 144$ frames. The machine’s physical constraints were set as follows: a peak gradient $G_\mathrm{max} = 40$ mT/m, a maximum slew rate $S_\mathrm{max} = 200$ T/m/s, and a sampling time $dt = 10$ $\mu$sec. The mean squared error (MSE) was used as the loss function in equation (\ref{eq}), with $\lambda_{ref}$ set to 5.

\subsection{Comparative Experiments} 
We demonstrate that our method surpasses the performance of Multi-PILOT \cite{shor2023multi}, the current state-of-the-art in dynamic compressed sensing to the best of our knowledge, both with learned and handcrafted acquisition trajectories. We follow the quality metrics from \cite{shor2023multi}, specifically PSNR, VIF \cite{sheikh2006image}, and FSIM \cite{zhang2011fsim}, as these have been identified as among the most reliable in the context of medical imaging \cite{pambrun2015limitations, mason2019comparison}. Moreover, we show that, similar to Multi-PILOT, our reconstruction model benefits from learning acquisition trajectories compared to using handcrafted ones. To validate this, we trained our reconstruction model with two sets of handcrafted acquisition trajectories: temporally-constant radial and time-varying Golden-Angle Ratio (GAR). In all experiments, we used 16 shots (RF excitations).

\label{compres}
Table \ref{tab:example} presents the accuracy metrics for the reconstruction experiments discussed in Section \ref{exps}. Our proposed 3D attention-based pipeline demonstrates superior reconstruction accuracy for both learned and handcrafted acquisition trajectories. Additionally, under the same data and batch size conditions, a single epoch in Multi-PILOT takes approximately 13 minutes, whereas the proposed pipeline reduces this time to around 6 minutes. The results presented were achieved with 180 epochs, while Multi-PILOT's per-frame optimization approach required 315 epochs to reach reasonable convergence. This indicates that the proposed method also provides a significant speed-up in training times. A visualization of the learned trajectories can be found in Appendix \ref{apptraj}.

\begin{table*}

\centering
\begin{tabular}{|c|c|c|c|c|c|c|}
\hline
\multirow{2}{*}{Acquisition Scheme}&\multicolumn{3}{|c|}{Multi-PILOT}&\multicolumn{3}{|c|}{TEAM-PILOT}\\
\cline{2-7}
& PSNR & VIF &  FSIM & PSNR & VIF &  FSIM\\
\hline
Const. Radial &$35.87\pm 0.74$ & $0.699\pm0.015$ & $0.8554\pm0.006$ &
$36.83$ $\pm$ $0.728$ &  $0.715\pm0.016$ & $0.8716\pm0.004$\\
\hline
Const. GAR & $34.30 \pm 0.61$  & $0.772\pm0.011$ & $0.822\pm0.009$
& $37.86 \pm 0.714$ & $0.812 \pm  0.014$ & $0.875 \pm 0.007$\\
\hline
Learned Pre-Ref. & $38.72$ $\pm$  $0.77$ & $0.823$ $\pm$ $0.009$ & $0.906 \pm 0.006$ &  \textbf{40.51 $\pm$ 0.79} & \textbf{0.83 $\pm$ 0.01} & \textbf{0.92 $\pm$ 0.005}\\
\hline
Learned Post-Ref. & - & - &  - & $40.35 \pm 0.80$ & $0.82 \pm 0.01$ & \textbf{0.92 $\pm$ 0.005}\\
\hline
\end{tabular}
\caption{\textbf{Reconstruction Results Comparison} - \textit{Learned} indicates trajectory learning. Pre-Ref. indicates results prior to trajectory refinement stage and Post-Ref. indicates full results.}

\label{tab:example}
\end{table*}









\subsection{Acquisition Time Reduction}
We demonstrate the potential of our method in reducing MR acquisition times. Let $N_\mathrm{shots}$ denote the number of RF excitations used, where 512 frequency sampling points are modeled for each shot. Our results show that the proposed method achieves similar reconstruction performance to Multi-PILOT while using fewer shots, leading to greater savings in acquisition time.

\begin{figure*}[htbp]

\begin{center}

  \includegraphics[width=1\linewidth]{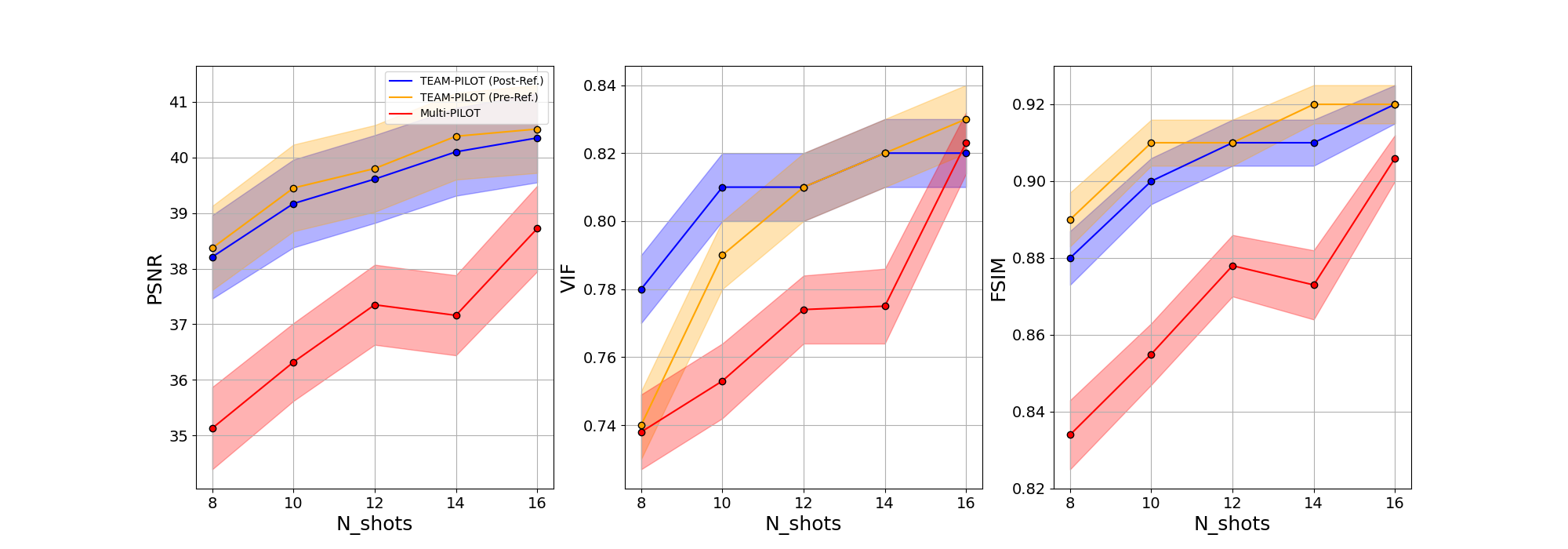}
  \end{center}
  {\caption{\textbf{Acquisition Time Minimization.} {Our results achieves results on-par with Multi-PILOT using 50\% less sampling points.} }
  
  \label{fig:nshots}
  }
\end{figure*}

\label{acqres} Figure \ref{fig:nshots} illustrates the reconstruction accuracy obtained with 8, 10, 12, 14, and 16 shots, across all evaluated metrics. Our method demonstrates strong reconstruction performance over a wide range of subsampling factors. Notably, TEAM-PILOT with 8 shots surpasses the performance of Multi-PILOT with 16 shots, indicating that the proposed method can achieve similar reconstruction quality with only half the samples used by Multi-PILOT, and with considerably shorter optimization times. Additionally, these results show that the trajectory refinement stage does not significantly impact reconstruction performance, even under more compressed sampling conditions (as compared to Section \ref{compres}).
  
\subsection{Temporal generalizability}
We provide both quantitative and qualitative evidence showing that our trajectory refinement (Section \ref{period}) effectively mitigates the issue of artifacts that occur during transitions between successive trajectory sequences, while having a negligible impact on final reconstruction accuracy.
\\
\label{perres} 
Figure \ref{fig:nshots} demonstrates that adding trajectory refinement has minimal effect on the final reconstruction quality. To illustrate the occurrence of artifacts and the effectiveness of our solution, we evaluate two trained models on sequences of 16 and 24 frames from the test set. The first model was trained on 8-frame sequences without trajectory refinement, while the second model was trained on 8-frame sequences with refinement ($\lambda_{ref}=5$).

\begin{figure}[htbp]
\centering

  \includegraphics[width=1\linewidth]{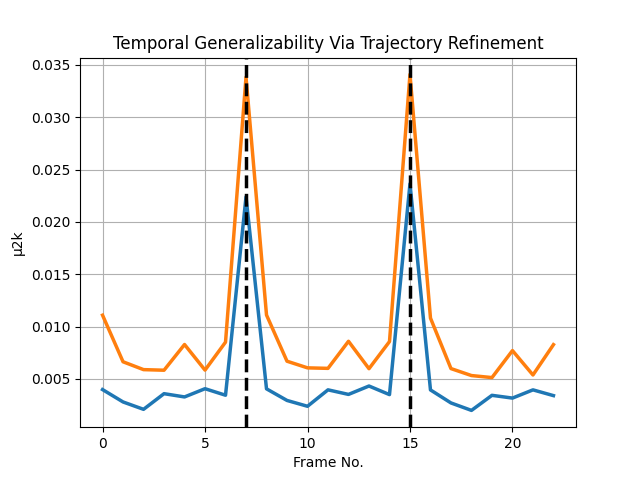}
  
  {\caption{\textbf{Mean Temporal Derivative $\mu_{2k}$ -} with (blue) and without (orange) trajectory refinement.}
  \label{fig:reg}
  }

\end{figure}
Figure \ref{fig:reg} shows the mean temporal derivative $\mu_{2k}$ (equation \ref{eq:min2}) with and without incorporating trajectory refinement. Figure depicts the artifacts mentioned in section \ref{period} in the form of large jumps in transitions between frames 7-8 and frames 15-16 (denoted by dotted vertical lines). As seem from the figure, while our method does not completely solve the rapid change between these subsequent frame pairs, it significantly diminishes it - by a factor of approximately 33\%. We provide similar plots for varying numbers of sampled points (including for sequence length 16) in appendix \ref{appref}.

\begin{figure*}[h]

\centering

  \includegraphics[width=0.65\linewidth]{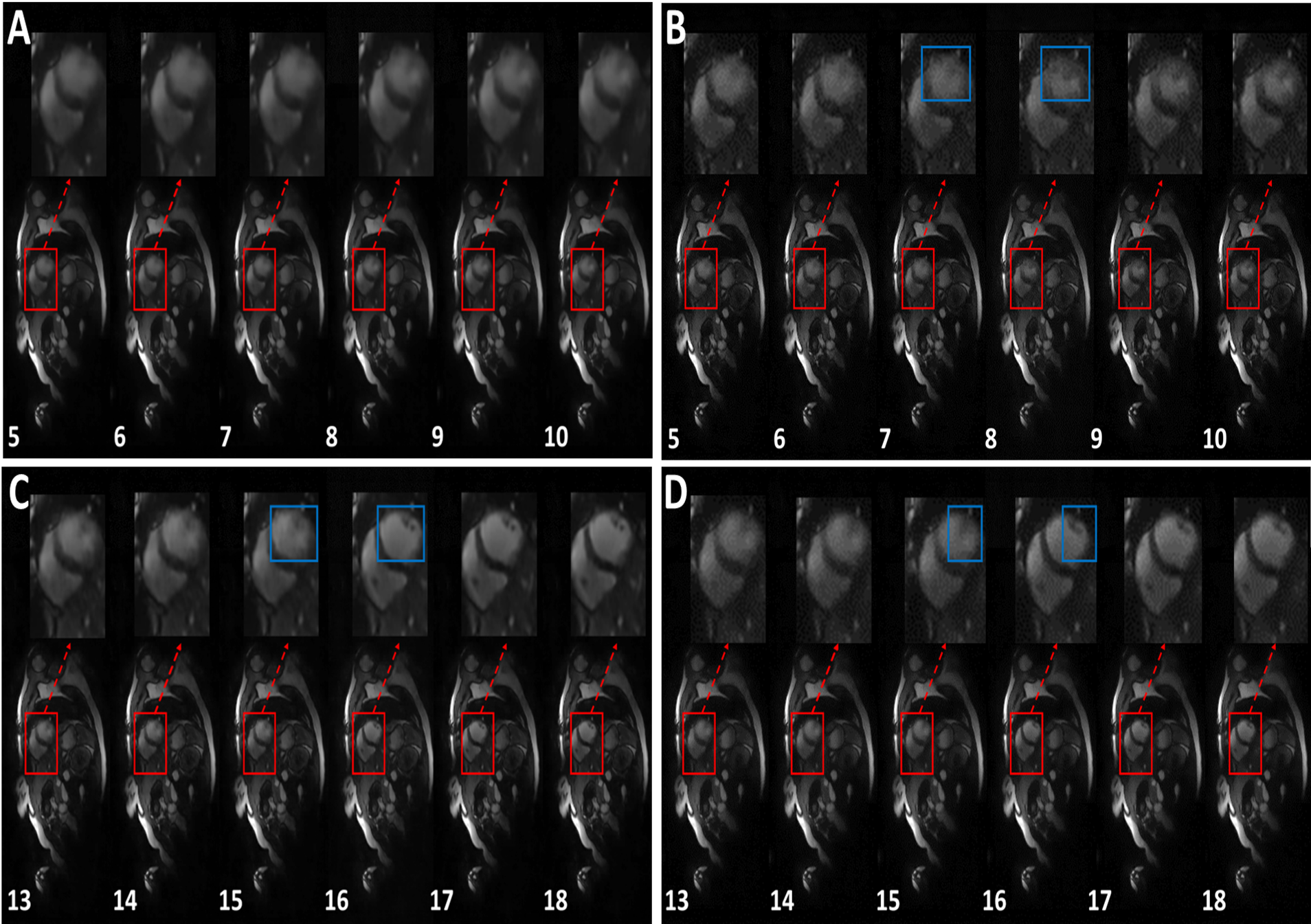}
  
  {\caption{\textbf{Reconstruction Results In Sequence Transition And Intermediate Frames -} {For inference of a 24 length sequence with trajectory stacking applied before (B,D) and after (A,C) trajectory refinement. Frame indices are in the bottom left.}
  \label{fig:reg_qual}
  }}

\end{figure*}

In Figure \ref{fig:reg_qual}, we qualitatively illustrate the conclusions drawn from Figure \ref{fig:reg}, presenting reconstruction results for the same 24-frame sequence around key frames of interest. Since we utilize trajectory stacking to perform inference on a 24-frame sequence using a model trained on 8-frame sequences, the transition frames between acquisition sequences are 7-8 and 15-16. \ref{fig:reg_qual}.A and \ref{fig:reg_qual}.C show results for reconstruction after the trajectory refinement stage for frames 5-10 and 13-18, respectively. \ref{fig:reg_qual}.B  and \ref{fig:reg_qual}.D are similar. However, they depict results before trajectory refinement. To demonstrate the artifacts mentioned in section \ref{period}, for each transition pair we also provide reconstruction results for two frames before (5,6 in A,B, 13,14 in \ref{fig:reg_qual}.C,\ref{fig:reg_qual}.D) and two frames after (9,10 in \ref{fig:reg_qual}.A,\ref{fig:reg_qual}.B, 17,18 in \ref{fig:reg_qual}.C,\ref{fig:reg_qual}.D) the frames of interest. These are a "control group" aimed to show that besides the transition frames, imaging of the heart motion is rather smooth in time - the major artifacts occur when transitioning between sequence groups. \\
In \ref{fig:reg_qual}.B we can see the mentioned artifacts occurring between transition frames 7,8 (highlighted in blue squares) - the upper-right region of the heart has a group of pixels abruptly becoming darker. The spatial smoothness of the image also abruptly changes between these two frames. In \ref{fig:reg_qual}.A (after refinement), however, this phenomenon does not occur. The transition between frames 7,8 appears smooth and similar to transitions between other frames (in the extent of change between subsequent frames). \\
In figure \ref{fig:reg_qual} \ref{fig:reg_qual}.C,\ref{fig:reg_qual}.D we show a case where our refinement method does not manage to fully eliminate the appearance of artifacts. Transition between frames 15,16 still has some jump in gray levels and sharpness compared to other neighboring frames (highlighted again in blue squares). Nonetheless, our method does diminish the extent of the phenomenon, as the artifacts appearing in transition between frames 15,16 in the pre-refinement case (\ref{fig:reg_qual}.D) seem more severe.

\subsection{Attention Maps Analysis}
\label{attres}
\begin{figure*}[htbp]

\begin{center}

  \includegraphics[width=0.8\linewidth]{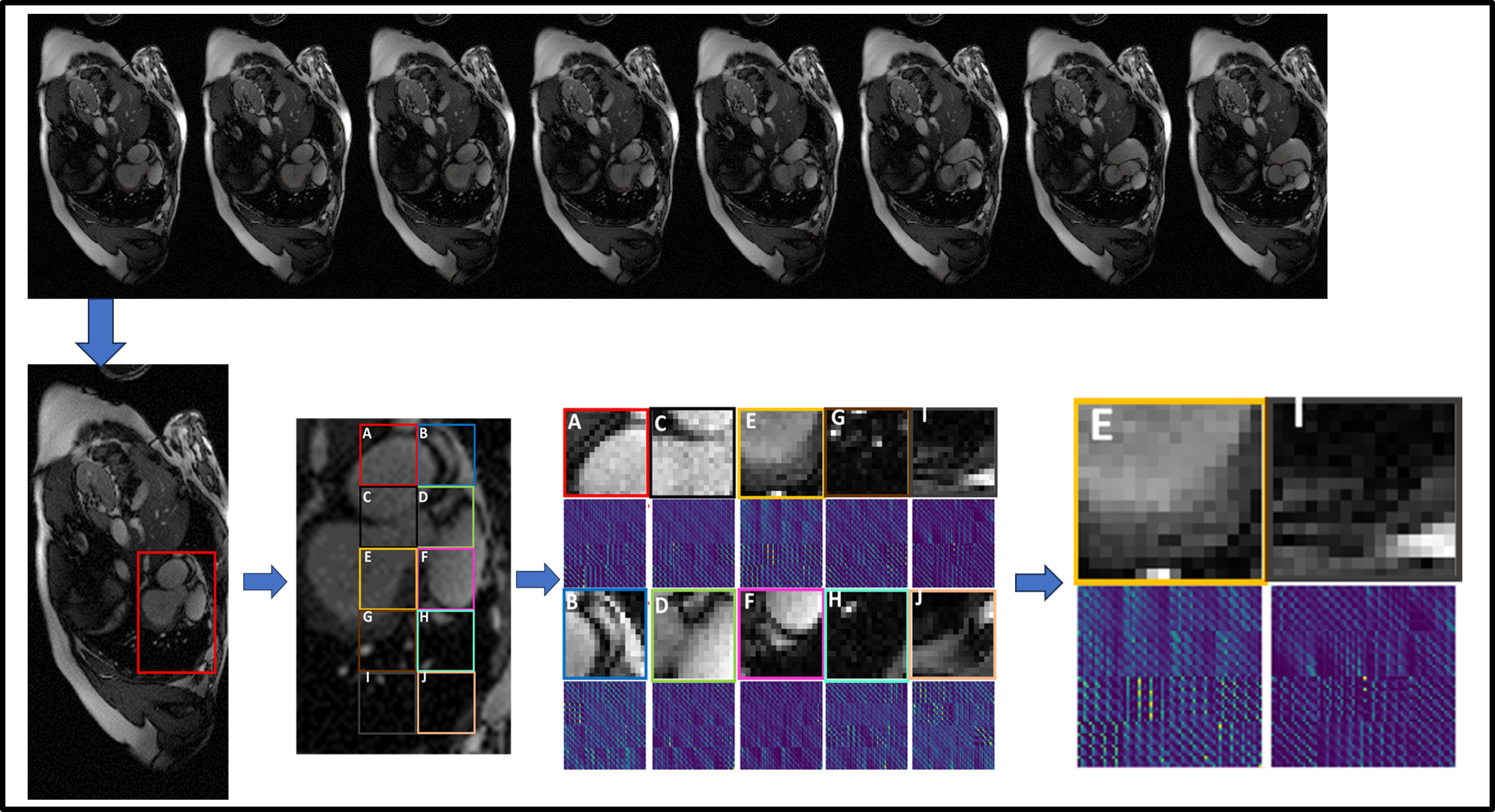}
  \end{center}
  {\caption{\textbf{Trained Model Attention Map Analysis - } {static patches receive higher attention scores along the main diagonal, while dynamic regions demonstrate wider spatial and temporal relations.} }
  \label{figat}
  }
\end{figure*}

In Figure \ref{figat}, we display attention maps for a dynamic region within a data sample. The upper row shows 8 frames from a specific video sample along the temporal dimension, where the chosen region of interest (marked by a red rectangle in the bottom left) is the most dynamic area—the heart. From this region, we selected 10 smaller regions, labeled \textit{A-J}, and present the resulting windowed attention maps below each selected image segment (second diagram from the right in the bottom row). For clarity, a focused view of the 10 attention maps can be found in Appendix \ref{app}.

In our model, attention windows are sized $4 \times 4$, producing attention maps with dimensions $16 \times 16$ per window. Each of the 10 selected image segments \textit{A-J} has dimensions of $16 \times 16$, resulting in 16 attention maps per segment (each attention map also being $16 \times 16$ in size). Each attention map corresponds to a specific $4 \times 4$ patch within the image segment. This relationship is further explained in Appendix \ref{app}. For the selected $16 \times 16$ image segment labeled E, the red squares below each represent 16 attention windows (with an enlarged view of this window array shown on the left). Each attention window corresponds to a $4 \times 4$ image patch, highlighted by the yellow squares under the diagram for image patch I (an enlarged view of a single attention map is on the right side of the figure).\\

As shown in Figure 5, our attention maps exhibit a trend correlated with motion. In more static regions (G-I, as seen in the upper row), the attention maps reflect a "static" pattern—where the highest attention weights align along the main diagonal of each $16 \times 16$ window (as exemplified by the enlarged diagram for patch I in the bottom right of Figure \ref{figat}). In contrast, in more dynamic areas (A-F), as demonstrated by image patch E, we observe larger deviations from the main diagonal across the windows, indicating that the model aggregates tokens from different locations to reconstruct the patch.

\section{Discussion}
\subsection{Conclusion}
In this work, we introduced TEAM-PILOT, a novel algorithm for Non-Cartesian Compressed Sensing of dynamic MRI. Our algorithm leverages a more efficient 3D attention mechanism to enhance current solutions, resulting in a performance gain of approximately 1.5 dB in PSNR, while requiring only about one-third of the training time compared to our baseline. We also highlighted the challenges of applying trajectory stacking for temporal generalizability and addressed these challenges by proposing a regularized trajectory refinement stage as an initial solution. This straightforward approach reduced visible artifacts by roughly 40\%. We validated our method by comparing it to both learned and non-learned dynamic CS pipelines and reinforced the findings from \cite{shor2023multi}, which suggest that learning non-Cartesian k-space acquisition trajectories leads to superior reconstruction results compared to non-learned acquisition schemes.

\subsection{Limitations and Future Work} 
Despite the promising results, we recognize certain limitations in our approach that we aim to address in future work. First, while our trajectory refinement reduces the jittering effect caused by trajectory stacking, temporal generalizability remains somewhat limited in our method. Second, our approach assumes that a dynamic MRI video can be acquired in discrete time frames. Although this assumption aligns with the format of the dataset used, it does not fully reflect real-world MRI acquisition processes. To enhance the clinical applicability of our method, more flexible modeling of acquisition trajectories will be necessary.

\printbibliography

\appendices
\begin{figure}[htbp]


  \includegraphics[width=1\linewidth]{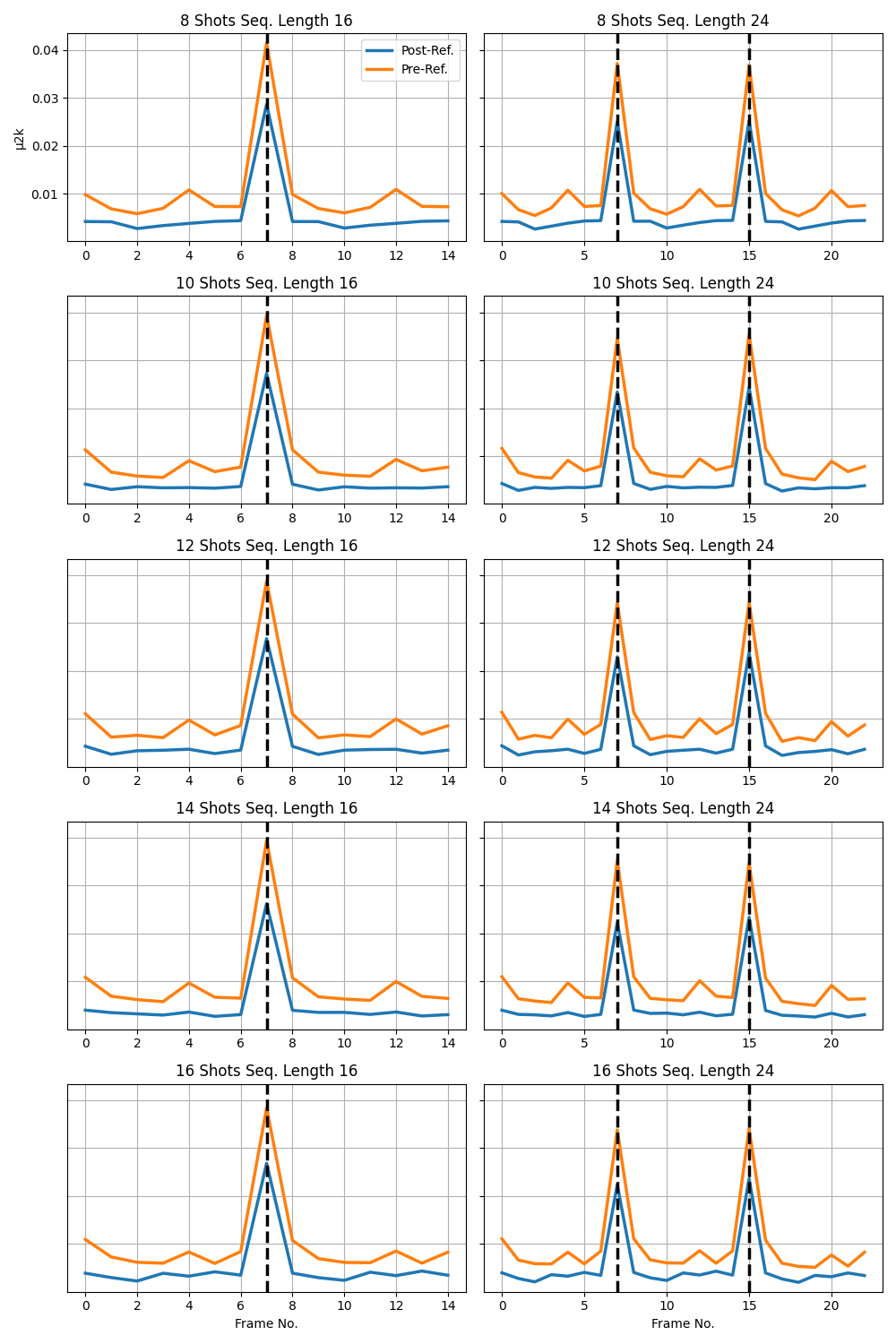}

  {\caption{\textbf{Mean Temporal Derivative $\mu_{2k}$} - for sequence lengths 16,24 and varying shots numbers. }
  \label{allregs}
  }
  
\end{figure}
\section{Learned Trajectories}
\label{apptraj}
\begin{figure*}[h]

\centering

  \includegraphics[width=0.9\linewidth]{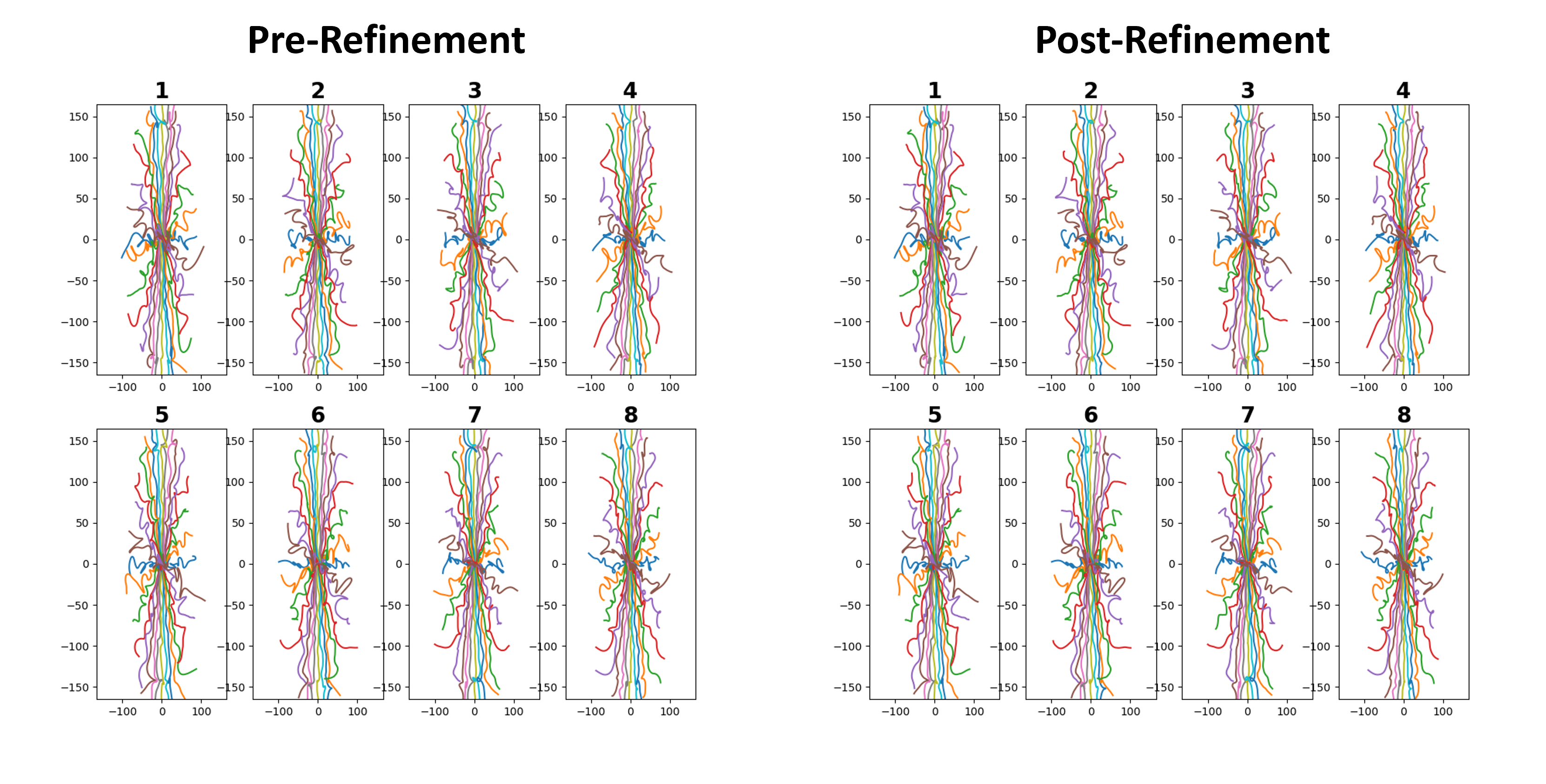}

  {\caption{\textbf{Learned Acquisition Trajectories - } {both after initial training and trajectory refinement.} }
  \label{figwref}
  }
  
\end{figure*}

In figure \ref{figwref} we show the per-frame acquisition trajectories learned by TEAM-PILOT both after the initial 170 training epochs (left) and after trajectory refinement (right). Axes represent k-space sampling coordinates. As expected, the two sets of trajectories are nearly identical, where there are only slight differences in the border frames 1 and 8, "connecting" one sequence to the other upon stacking. \\
Regardless of trajectory refinement, similar to Multi-PILOT, we can see that trajectories in earlier frames are more concentrated on the central vertical axis. In later frames trajectories seem to fan-out more onto wider vertical axes, indicating efficient data-transfer between different frames, allowing later frames to focus on new information in higher frequencies, after the core static information had been captured by the initial trajectories.

\section{Temporal Generalizability}
\label{appref}
In figure \ref{allregs} we show the mean temporal derivative plot (figure \ref{fig:reg}) for trajectory stacking evaluated with sequences of lengths 16 and 24, across varying numbers of shots. Transition frames 7 and 15 are again highlighted in dotted lines. The overall trend from figure \ref{fig:reg} is preserved - we see peaks upon the transition between sequences, and our method of trajectory refinement manages to reduce this effect by approximately 33\%. It also seems our method becomes less efficient as the number of sampling points decreases - we attribute this to the fact that when using fewer sampling points, we have fewer degrees of freedom when shaping trajectories to both produce good reconstruction and temporal smoothness.

\section{Attention Maps Analysis}
\label{app}

\begin{figure*}


\centering
  
  \includegraphics[width=0.8\linewidth]{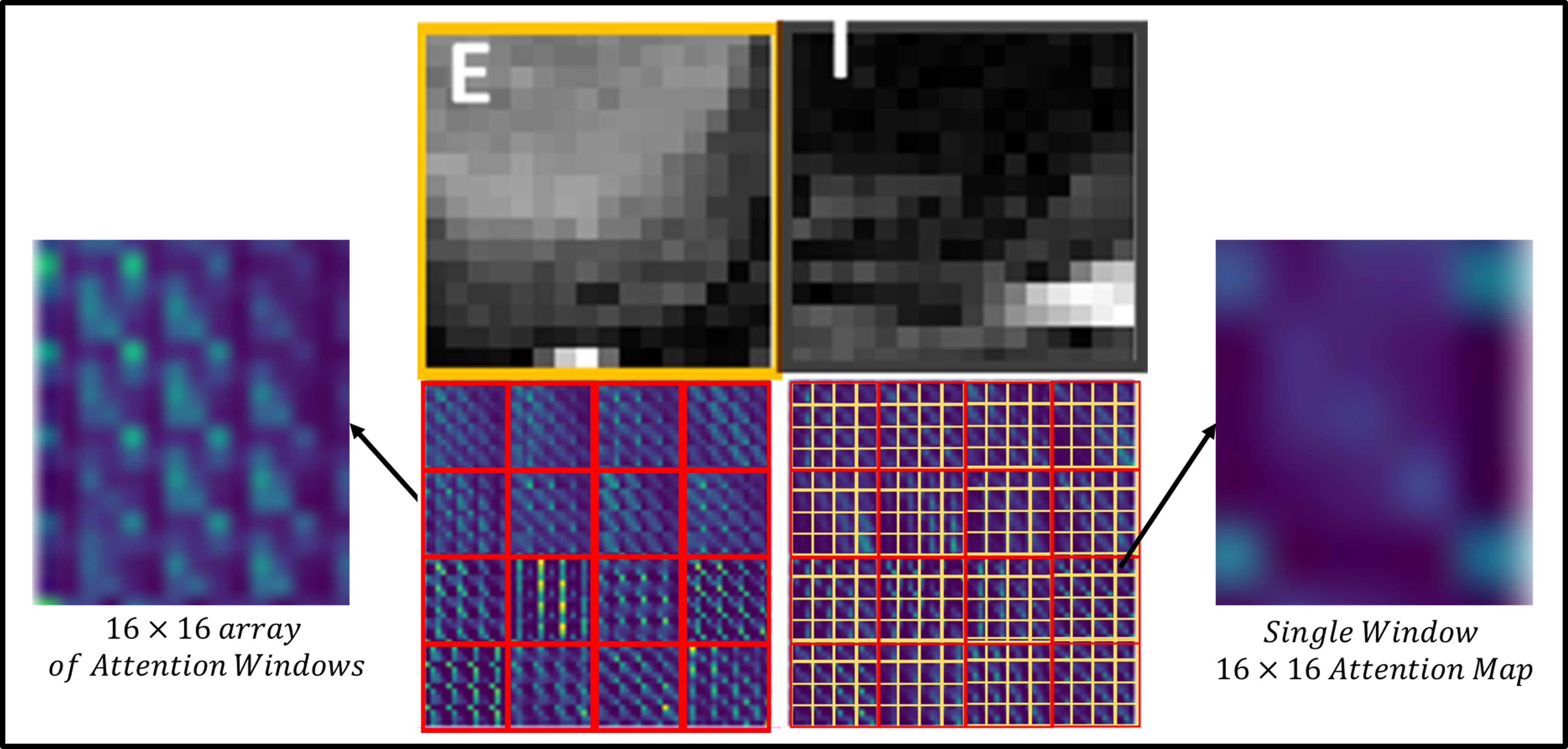}

  {\caption{\textbf{Image patch partitioning under window attention - } {Each $16\times16$ patch (red) contains 16 attention windows, each of dimension $16\times16$ on their own (yellow).} }
  \label{figw}
  }
\end{figure*}

Figure \ref{figw} further demonstrates what is seen in the visualization shown in section \ref{figat}. The attention windows are in red (bottom left). Within each window is a $16\times16$ attention map yellow). 

\begin{figure*}


\centering
  
  \includegraphics[width=0.8\linewidth]{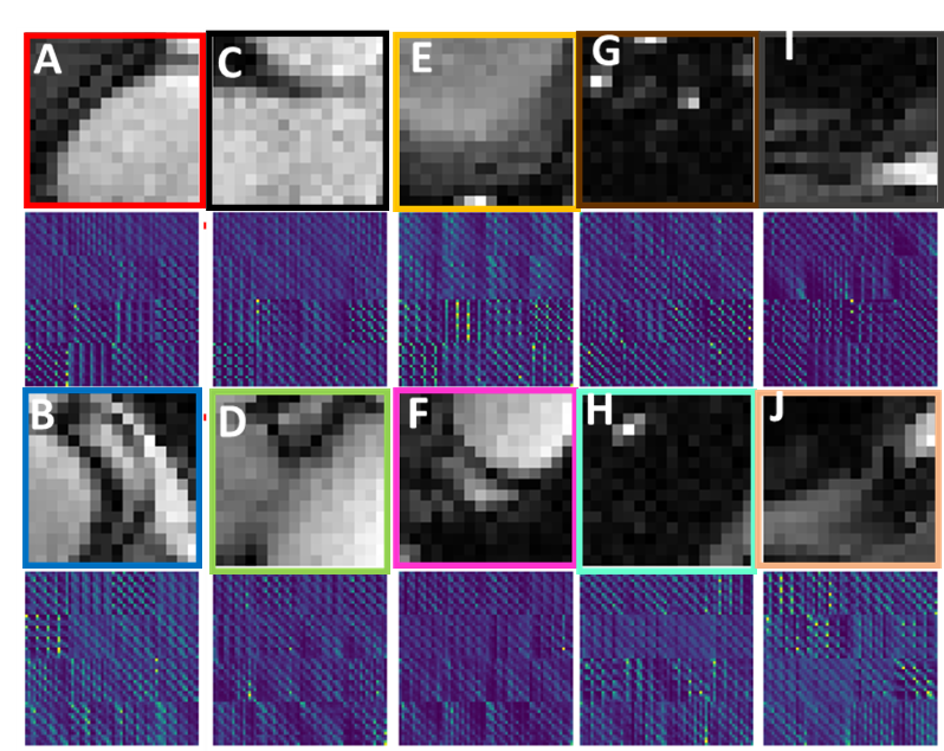}

  {\caption{\textbf{Focused view on attention maps from section \ref{figat}} }
  \label{figw2}
  }
\end{figure*}

\end{document}